# Reworking the SETI Paradox: METI's Place on the Continuum of Astrobiological Signaling

Thomas Cortellesi

*Department of Geophysical Sciences, University of Chicago, Chicago, IL, USA*

*Correspondence: Thomas Cortellesi <tcortellesi@uchicago.edu>*

**Keywords:** METI, SETI, Astrobiology, Technosignatures, Biosignatures

## Abstract

The Search for Extraterrestrial Intelligence (SETI) has heretofore been a largely passive exercise, reliant on the pursuit of technosignatures. Still, there are those that advocate a more active approach. Messaging Extraterrestrial Intelligence (METI) has had a controversial history within the larger SETI project; it is claimed that the risks involved outweigh any potential benefits. These arguments are ultimately not compelling, result in absurd policy recommendations, and rest on a faulty appreciation of the nature of technosignatures, whose detectability implies intent to signal. Present technology is advancing quickly such that we will soon have great observational reach, to the point of reliably detecting such technosignatures and biosignatures: a capability that can be matched or exceeded elsewhere. To escape the SETI Paradox properly defined, at least one technological civilization must choose not to suppress its own continuum of astrobiological signals, of which METI is merely the most effective endmember. Passive SETI's low likelihood of success in the short-term is a serious obstacle to sustainable funding, alongside a 'giggle factor' enhanced by a pernicious fear of contact. The scientific community must integrate an active approach to better ensure both the continuity and eventual success of the SETI project.

## 1.0 Introduction

The plurality of worlds is an ancient dream: thanks to modern astronomy, we understand it at last to be reality. So long as we dreamt of those worlds, we often supposed them inhabited, usually by beings quite like us. The Search for Extraterrestrial Intelligence (SETI) has spent the last six decades scanning the skies for signs of such extraterrestrial technologists (ETs). Despite plausible estimates that project dozens of other technologically advanced civilizations to exist within the galaxy [1], to date none have been found. This is the Great Silence: a puzzle that has since its inception captivated the astrobiological community, and many others besides. Where are the aliens?

There are two established ways to answer this question. One is to detect a biosignature: some 'object, substance and/or pattern whose origin specifically requires a biological agent' [2]. On worlds presently unreachable by in-situ analysis and remote sensing by spacecraft, these phenomena will be rendered as one or more spectral features that indicate atmospheric gases or surficial reflection, best considered in the context of their planetary and stellar environments [3,4].

Some traditional biosignatures include $O_2$, $O_3$, nitrogen oxides (NOx), complex sulfur gases, photosynthetic pigments, redox disequilibria, or seasonal oscillations of the same [3,4].

SETI sets itself apart from the larger astrobiology project as the hunt for technosignatures: detectable features indicative of an environment modified by technology [5]. Common biosignatures can be created by abiotic processes, but technosignatures provide increased confidence in the detection of extraterrestrial life [3]. Many kinds of technosignatures have been proposed, and different SETI disciplines have evolved to find them [6]. Communications SETI aims to find communications leakage or directed messaging from ETs across the electromagnetic spectrum. Artifact SETI searches for ET's environmental footprint, manifested such things as artificial structures, atmospheric pollution, waste heat, or space probes a-la-*Voyager*. Inexplicable anomalies in nature can also point to the presence of ETs, though with less certainty. Traditionally, most time and effort has been dedicated to radio SETI, as immortalized in Carl Sagan's *Contact*.

The object of such 'passive' SETI efforts is to search for technosignatures across the electromagnetic spectrum. Still, there are those within the community that advocate a more active approach. METI (Messaging Extraterrestrial Intelligence), or Active SETI, involves the intentional sending of messages or recoverable probes, in the hopes of eliciting a reply. To date, 16 distinct messages have been transmitted to 26 different targets [7], plus the *Voyager* Golden Records and *Pioneer* plaques. Taken liberally, the five spacecraft humanity has sent on interstellar trajectories themselves count as METI. Despite this storied history, METI's position within the larger SETI project has long been controversial. It is often seen, quite literally, as 'searching for trouble' [8]. The argument over METI's legitimacy and effectiveness has raged since the Arecibo Message was sent in 1974 and continues to this day.

### 1.1 The METI Debate

METI proponents, hereafter *metiists*, have historically proffered many arguments in favor of their position. Chief among them is advocacy for a diversity of strategy in the SETI project, especially after so many years of non-detection. Doug Vakoch, head of METI International, maintains that 'through METI we can empirically test the hypothesis that transmitting an intentional signal will elicit a reply' and that objections to METI suffer from various psychological biases [9]. He and Shostak claim that any ET with somewhat greater technical capacity could detect humankind's radio communications leakage [9,10], and Zaitsev takes this a step further by declaring this makes it impossible to hide from more advanced civilizations [11]. Additionally, powerful radar soundings of asteroid and planetary surfaces have proved integral to the study of their morphology and dynamics, and treating such transmissions as dangerous could discourage dedicated systems from being built, and reduce humankind's ability to precisely chart the ephemerides of Potentially Hazardous Objects [12]. More prosaically, Cold-War era over-the-horizon radars were powerful and even diurnal radio sources that could be detected by ETs [13].

In short, metiists typically maintain that our communicative, technological society has betrayed our presence to ETs, whatever their disposition; though Vakoch maintains that SETI itself relies on altruistic messaging from the other side [9]. Metiists tend to hold that ETs are likely neutral or benign, and paint rosy pictures of the outcomes of contact. They offer humanistic reasons to engage in METI too – in distilling what is best about humankind, singular low-impact efforts like the Arecibo message and *Voyager* Golden Records stand far more robustly as messages *to humans* about SETI than as practical attempts to contact aliens. Transmissions also act as time capsules to provide information about civilizations that may no longer exist, and consequently have archaeological and cultural value [14]. Zaitsev provides perhaps the most heartfelt arguments, looking to history to say that *not* reaching out to other civilizations could lead to the human species' extinction through apathy and isolationism. Most poetically, he frames METI as an answer to the Great Silence [9].

METI opponents, hereafter *passivists*, have developed answers to these arguments and claims of their own. They maintain that passive SETI's lack of success so far is no indication of its capacity to accomplish its goals and that tempting fate by transmitting constitutes 'reckless endangerment of all mankind' [15,16]. Whether it is asserted that we cannot know the motives of ETs [8], and ought thus beware, or that we can, and they must be accordingly malign; whether it is imagined technologists would travel to Earth to harvest our resources [17], enslave humanity [18] or, more insidiously, have our world destroyed remotely with relativistic or Drexlerian weapons [15], the spectrum of possibility is deemed too great to hazard.

METI advocates stand accused of 'following a naïve faith' that ETs are altruistic or godlike, maintaining breathlessly that the aliens are waiting for us to call, eight arms outstretched in welcome to the Galactic Club [15]. Indeed, some metiists *are* partisans of the Zoo hypothesis, and optimistic to a fault. Without concrete plans to receive a return message, metiists are said to have no methodology nor sense of empiricism. Indeed, they are hardly scientists at all, more comparable, as passivist-extraordinaire John Gertz acidly puts it, to 'someone who cultures and releases anthrax' [16].

Evolutionary psychologists have attempted to condemn METI as announcing Earth's myriad resources to the cosmos while leaving it 'lightly guarded by a gullible young species' [17]. Similarly, luminaries like Stephen Hawking have compared METI to inviting European colonizers onto First Nations soil; he maintains that contact had better wait until humans are further developed as a species, both technologically and ethically [19]. The risks of METI today, it is plausibly argued, outweigh the benefits – humankind should instead listen passively for decades, centuries even, before attempting to initiate interstellar contact. Gertz also calls on governmental bodies to make an injunction against METI and prohibit intentional signaling to extrasolar targets at power greater than a terrestrial television broadcast [16].

That specification is considered important, because passivists maintain that humanity's communications leakage cannot be detected by ETs [20]. More to the point, they say METI will not be detected, or if it were, it will not be decoded [13]. Even the 'best case scenarios' (e.g. beams of microwaves used for power transmission) would appear to ETs only as transient events. Moreover, passivists maintain that Earth's detectability window is closing; communications will soon take place through cables, or over directed satellite links [8,15]. Why run the risk when we have yet to show our hand?

Passivists claim too that funding is in the bag and that passive SETI's progress is due to accelerate. The *Breakthrough Listen* program recently brought in $100 million over 10 years from magnate Yuri Milner, and passivists duly expect that public funding will follow [8]. Most interestingly, some maintain that the search is nearly over, and ET artifacts could already be in the Solar System, waiting to be discovered [15]; only time will tell.

## 2.0 Discussion

It is certainly true, as admirably illustrated by passivists, that many pro-METI arguments fall flat. There is no guarantee that ETs will pick up our communications leakage by itself, or understand its significance, or be able to understand it. The zoo hypothesis is contrived; SETI has indeed just begun its search, and at any rate does not rely exclusively on directed, altruistic messaging. However, little else in the passivist arsenal bears scrutiny, and this debate has been limited almost exclusively to 'orthodox' radio-SETI [6]. As such, passivists neatly skirt the crux of the issue: they fail to see METI's scientific potential and humans as both participants in and objects of the Search for Extraterrestrial Intelligence. Even so, it is still worthwhile to answer detractors' points specifically before moving on to the larger picture.

### 2.1 Challenges to Passivist Arguments

Trivially, both EM leakage and METI *can* be detected by ETs with sufficiently powerful telescopes; pointing such fantastic instruments in the right direction is a separate if soluble problem, as will be shown later. Moreover, even if transmissions cannot be interpreted, they can provide evidence of a technological civilization in the vicinity of Earth [14], which may be enough to elicit a reply (or, taking passivists at their word, hostilities). Assuming targets to be incapable of decoding transmissions is liable to underestimate the tenacity of advanced species. It is true that metiists are often guilty of using numinous language to describe such ETs; Zaitsev refers to 'highly developed civilizations' as 'something supernatural and mysterious' [11]. This, however, is Clarkean aphorism, not a religious profession – indeed, Gertz makes the same point about magiclike extraterrestrial technology *ad nauseam* [16]. If all that is left to the charge of 'faith' is optimism, then so be it: some humans are condemned to be sanguine.

This does not entail, however, that metiists assume anything about ET's behavior. Indeed, passivists cannot seem to agree on whether *they* can 'know the mind of extraterrestrials' and predict the outcome of contact, the source of the characteristic risk they perceive. Neutral outcomes to contact are just as possible as positive or negative ones [21], entailing that METI will still be useful in all cases save one where we stumble on a 'hostile super-civilization' [18].

Positing that the response of an extraterrestrial intelligence to contact would be malign or predatory relies on assumptions that *cannot be substantiated until contact occurs* [22]. Such projections make themselves predicate on one's priors, and on analogies to human culture [23]. Tempting though it may be to predict alien behavior by looking to human evolution and history, it seems dubious to extrapolate too much from a sample size of one; it is impossible to know whether (aggressive) human behavior is characteristic of intelligent life, though serious attempts have been made to suggest that this is the case [17]. Granted, there is no reason to positively expect a Federation of United Planets either; this is the stuff of science fiction, as are hysterical fears of alien invasion. Indeed, Gertz refers to his own cautionary hypotheticals as 'outlandish' [16], and for good reason. Any disinterested intelligence capable of travelling to Earth and vaporizing our oceans accordingly has the power to fix its own large-scale problems locally, with less effort – and that is provided that our existence is immaterial to it, which may not be the case. Life, or intelligence at any rate, may have value to ETs, or not. There *is* a persuasive element to the axiom that civilizations harmonious enough to last for eons do not, by definition, blow themselves up within the characteristic time frame, but again that is absolutely no guarantee of anything.

More concretely, the incentive for hostility between widely spaced ETs is misaligned. It may be possible to estimate an ET's scientific capabilities in relation to one's own, and perhaps their rate of technological progress and aptitude for aggression, but over vast, interstellar distances the margins of error in these calculations could range many orders of magnitude, a fact as true for remote weapons as crewed visitations. What if ETs, a thousand light years distant, were to receive one of our messages? Humanity could be gone by then, annihilated in a thermonuclear exchange – or, it could be busy settling the nearest twenty systems, playing with Death Stars and laser swords. Granted, there may be unknown methods to escape this time delay, to travel here instantly or know of humankind's state of affairs in real time – but then, who is pretending aliens are gods?

There *is* a kernel of truth to passivists' claim that METI efforts have been unscientific; however, these are problems of particulars and not fundamental to the METI framework. Certain past transmissions are better thought of as publicity stunts than serious attempts to contact ETs [7]. Additionally, Vakoch's formulation of METI's scientific potential is true but hopelessly narrow and ought be broadened as the following: *METI tests for ETs by actively inducing the data that orthodox SETI waits passively to receive, increasing sampling efficiency and accelerating data acquisition.* It is by design a different approach to the same problem (seeking to find ETs) and deserves its place within the larger, interdisciplinary SETI project [6]. While passivists maintain that astronomy is purely an observatory science and active approaches to it are incoherent, such a view ignores previous interdisciplinary efforts that did exactly that: e.g. the collision of the *Deep Impact* mission's impactor with Comet 9P/Tempel, which was observed both by the spacecraft bus and a number of other space-based and Earth-based telescopes.

Again, to passivists' credit, metiists have also not yet articulated a coherent plan to receive return messages; however, this is not entirely their fault. Passivists have determinedly thwarted attempts by metiists to obtain time on large, powerful radio telescopes and without that foot in the organizational door, planning for return messages has been made difficult. Moreover, much like the hostility argument, considering metiists' lack of return plan important fails to appreciate the time delay inherent in sending and receiving messages. While most METI and passive SETI efforts have been focused on nearby stars (scientists like to have questions answered within their lifetimes), many of the most interesting targets of astrobiological scrutiny are hundreds if not thousands of light-years away. Decades or more of time in the interim allows for construction of new astronomical infrastructure (some of which is underway now) and the plan to receive amounts itself to passive SETI. Critically, searching for return messages operates over a small area of sky and a restricted time interval; as such, arguing that we will miss such directed messages reduces to arguing that we will reliably miss unplanned-for transients (at any time and from any direction in the sky) otherwise interesting to orthodox SETI. METI is, in effect, working to speed up the process of contact, to settle the longstanding empirical question posed by the Great Silence once and for all.

Hawking sees this acceleration as a detriment, maintaining humanity needs time to develop the necessary ethics and advanced technology to smoothly engage with extraterrestrials. This is, however, a line of argument just as dangerous to the larger SETI project as to METI. Following his logic and thus supposing humankind were 'not ready' for contact, our own SETI programs would be pointless in the short-term: should we find anything, it would seem at present we are civilizationally incapable of dealing with it. This is an absurd conclusion to draw, as it is not obvious what the effects of discovery would be on the human psyche [24].

Hawking's position further voids the push to discard METI on the grounds that SETI will succeed in the near future. This often relies on the assertion that ET probes are here in the Solar System; much though one would like to live in the world of *2001: A Space Odyssey*, we cannot work on the assumption that we do, or that ET exists at all [25]. In that case, why not do what we can increase the likelihood of discovery?

Similarly, the assertion that Earth's EM detectability window is closing relies on uncertain projections of the future, not just of technology but of its placement within the Solar System. One can imagine scenarios where high-gain, high-power interplanetary communications may be necessary; targets will often be smaller than the cross-section formed by laser or radio beams. Even granting passivists' hypothetical, the assertion that Earth will soon be undetectable *is not and cannot be true* in the broadest sense.

## 2.2 Earth as a Planet: Astrobiology and the Copernican Principle

In proposing to regulate Earth's display to the cosmos, passivists consider only orthodox radio technosignatures – yet human activity leaves a planetary-scale, characteristic environmental footprint amenable to an Artifact SETI approach. Atmospheric contamination by artificial pollutants such as chlorofluorocarbons (CFCs) [26], industry-relevant particulates like heavy metals, and geologically rapid changes in trace but significant gas concentrations (CFCs, $CO_2$,

NOx, CH$_4$, O$_3$) can provide potential technosignatures. Industry has integrated itself with global biogeochemical cycling and that is only projected to intensify as time goes on [27]. As we brace to counter and contend the effects of climate change, it is worthwhile to note that Earth is in 'hybridized' thermodynamic disequilibrium, a state also considered a potential technosignature [28], though at present the contribution of direct waste heat from human civilization is small. City lights illuminating Earth's nightside, with their diurnal variance and artificial spectral signature, could also be detected by telescopes relying on well-understood technology [10]. It is easy to forget that to any putative aliens, *humans* are ETs too, and thus in considering Earth's signaling potential *all* technosignatures need to be accounted for – not merely the ones that are convenient.

Passive SETI relies on the assumption that some other civilization, having confronted the cosmic emptiness, must have already decided that being detectable to others was worth the risk it entailed; that is, that the detectability of technosignatures *conveys the implicit intent to signal*. At present, humankind is behaving as though detectability were not a problem, engaging neither in an active messaging campaign nor in serious repression of communications leakage or artifact technosignatures. This neutral approach illustrates how concerns about hostile ETs are not held seriously and constitutes unplanned signaling.

Less obviously, biosignatures are part of this conundrum, as they too are by definition detectable. Additionally, they serve as grounds for a targeted, follow-up search for technosignatures if identified. From this, a continuum of astrobiological signaling can be established, ranging from unintentional biosignatures, to passively broadcast technosignatures and onward to active messaging. METI thus stands as an extreme, directed, active technosignature, the difference between it and other astrobiological signaling being one of degree, not kind.

Spectral technosignatures and even biosignatures are on the edge of humankind's detection capabilities now; soon we will have great observational reach to the point of reliably identifying these biogenic phenomena at great distances. The long-delayed JWST will dependably identify atmospheric technosignatures on nearby planets after only a day or so of integration time [26]. The upcoming Nancy Grace Roman Space Telescope (formerly WFIRST, the Wide-Field InfraRed Survey Telescope) and Vera C. Rubin Observatory (formerly LSST, the Large Synoptic Survey Telescope) are both dedicated in part to the discovery of exoplanets [29,30]. Global projects such as the Square Kilometer Array will greatly enhance our radio astronomy capabilities, and LUVOIR (the Large UV/Optical/IR Surveyor) will launch with the intention to directly image and spectrally characterize the atmospheres of exoplanets [31,32]. In the last decade, the notional cost of a kilogram to orbit has dropped by an order of magnitude and may soon fall another thanks to innovative space ventures, enabling still more fantastic space-based telescopes in the coming decades. New light collection techniques taking advantage of the lensing effects of Earth's atmosphere or the Sun's gravitational field have also been proposed [33,10]. Armed with these extraordinary instruments, in the coming decades scientists will peer into the atmospheres of nearby exoplanets identified by wide-field surveys [34] and listen to the sky with still greater aptitude. Additionally, we are developing new and complimentary techniques in characterizing habitable (and, ultimately inhabited) worlds [35].

If, as passivists admit, other civilizations are likely much older (and thus liable to be much more advanced) than our own, will they not have still more capable telescopes and techniques, and perhaps an interest in using them? Even well-understood technology, scaled to sizes soon achievable, enables remarkable observational reach without resorting to Clarkean maxims. Again, SETI is a two-way street, and passivists fail to apply the mediocrity principle and see humans as extraterrestrial technologists. Theirs is an anti-Copernican viewpoint, arrogating to humans a special reference frame where other ETs' astrobiological signals are detectable, but ours are not. This is spurious. Exoplanets represent natural 'focal points' in the search for technosignatures [36]; Earth is thus liable to be a target of astrobiological scrutiny in the same way other 'potentially habitable' exoplanets have been by humans [37]. Worlds considered habitable that are also bizarre sources of transients (e.g. radar bursts, radio chatter, rapid atmospheric spectral changes, etc.) could be determined to be of special interest, and potential technosignatures that are suspect alone can be made convincing together.

Life has modified Earth's environment considerably throughout its ~4 billion-year tenure on the planet and modeling the detectability of these changes throughout Earth's past (hoping to inform studies of Earthlike exoplanets) has become something of a cottage industry. $CH_4$ was a thousand times more abundant in the Archean than today; it and photochemical organic hazes derived from it represent two such biosignatures [38]. While the Great Oxidation Event (2.5 Gya) knocked $CH_4$ below plausible detection thresholds and $O_2$ may have been detectable only for the last ~10% of Earth's history, $O_3$ discernable through strong ultraviolet features provides a strong proxy for biogenic $O_2$ since the Paleoproterozoic, stopping the gap [39]. If Earth has been displaying biosignatures continuously for the past 4Gyrs, then it may be detectable by ETs even at extraordinary distances and through deep time.

In this light METI does not present a unique risk to humankind; those preexisting 'hostile super-civilizations' dedicated to snuffing out other evolutionary experiments in intelligence will have already had considerable time to discover Earth's biotic potential and act accordingly; if they were nearby and wished to find us, they surely could. While some metiists might contend that only intelligent life (or perhaps only those species actively wrecking their planetary ecosystem) would be a cause for aggression, again biosignatures stand as a reliable precursor to technological societies that could provoke such hostility. If these advanced societies are old, as they are likely to be, then it stands to reason that Earth may already be in their proverbial crosshairs.

**2.3 Reworking the SETI Paradox**

These facts present a dilemma known as the SETI Paradox. First coined by Zaitsev, it is traditionally phrased as 'SETI… is meaningless if no one feels the need to transmit' [40]. It can also be rendered as a question: 'what is the point of SETI if transmission is not worth the risk? We will never find anything!' In this wording, it is specifically unsymmetric 'transmission' that leads to failure of discovery; unfortunately, like Vakoch's formulation of METI's empirical practice, this definition is contrived and narrow. Additionally, it limits itself to the sphere of

Communications SETI and relies to some degree on the false claim that passive SETI depends on altruistic messaging. Thankfully, this is not the whole story.

Communications SETI has been likened to a cooperative game in which players cannot communicate [6]. Framing the SETI Paradox exclusively in terms of transmission buys into the passivist misapprehension that Earth is safe from detection if no interstellar transmissions are sent. The broader view of detectability established herein incorporates METI onto a signaling continuum that includes passively displayed biosignatures and technosignatures, both of which are broadcast merely by our presence on Earth. This makes the larger SETI project an opt-out, not an opt-in game – voiding the traditional SETI Paradox, or at least requiring it to be reframed in proper terms.

If civilizations truly are risk-averse, they will not just avoid METI programs, but also minimize communication leakage, and perhaps not engage in large-scale activity of the kind that would be visible to other ETs. Inordinately cautious ones could suppress atmospheric and surficial signals too, for fear of being found [22]. If all nearby civilizations do this, then none of them will discover each other. This is the reworked SETI Paradox: *SETI is meaningless unless at least one alien civilization chooses not to suppress its own panoply of astrobiological signatures.*

This restores SETI to its base assumptions (that civilizations are detectable both through technosignatures and biosignatures) while treating humans as what they are: ETs that can be detected in the diversity of ways we ourselves search for them. A rephrased corollary follows: what is the point of SETI if *signaling* is not worth the risk? SETI predicates itself on the idea that civilizations will exhibit detectable technosignatures, and life biosignatures (though not necessarily at the same time or in the same place – risk-averse civilizations may transmit from locations far from their homeworld); if any these signals are not forthcoming, then SETI is pointless.

METI's place as the endmember on Earth's signaling continuum underlies the fundamental incoherence to any pro-SETI, anti-METI position: if no one signals, no one receives; if METI is unscientific, then SETI is unscientific; and if we aren't ready to initiate contact, we certainly aren't ready to receive messages to that effect. Every single viable case against METI reduces to one against SETI: a practice that, while not without controversy, is a well-established scientific discipline. Passivists' lack of imagination in this respect and selective application of the Copernican Principle entails, as will be seen, ludicrous and irresponsible policy recommendations.

### 2.4 Passivists' Absurd Policy Recommendations

Passivists claim that METI presents a unique, existential risk to humankind and further claim that legal countermeasures ought to be taken against it. Gertz gleefully proposes that transmitters be 'proscribed with criminal consequences' on the national or even international level [16], conjuring up fanciful images of SWAT teams descending on the homes of unassuming ham radio operators. More sweepingly, he proposes to prohibit intentional signaling to extrasolar targets at power greater than a terrestrial TV broadcast. This harkens back to the passivist assertion that such signals are not detectable, which can only be considered reasonable when evaluated simplistically; remember, the numerous technosignatures and biosignatures Earth exhibits alongside its special place in the LWHZ (Liquid Water Habitable Zone) cannot be considered separately. Beyond being arbitrary, this injunction is entirely unenforceable [41], and has several downstream consequences.

Metiists have expressed fear that such prohibitions would interfere with asteroid and planetary radar soundings [12], and targets of powerful EM transmissions will inevitably be smaller than the beam's distant cross-sections themselves. Gertz anticipates this objection by proposing that such transmissions should be prohibited only when targets occult background stars within 1000 LY [16]. Again, this cutoff is arbitrary and fails to appreciate Earth's long and continuous history of astrobiological signaling. The idea seems workable in principle – but see how willing passivists are to squander and complicate valuable telescope time. Additionally, does this propose to limit and interfere with our descendant's ability to communicate across interplanetary and even interstellar distances [10]? Does this entail a moratorium on lightsail-type interstellar probes, given the massive laser arrays (aimed at or near nearby stars) needed to accelerate them to relativistic speeds? This attitude seems strange, given passivists' enthusiasm for the other doings of *Breakthrough Initiatives*. Lastly, as Gertz weirdly and straw-clutchingly puts it, METI specifically endangers American national security by transmitting information to ETs who are not in Earth's public domain [16]. Take note of appeal to militarism; it will come back.

Conveniently, Gertz' and other passivists' policy proposals entail no serious change to the status quo. However, as they are loathe to realize, radio communications are not Earth's only brand of technosignature. Haqq-Misra terms passivists' policy position as 'precautionary malevolence' and correctly identifies the next step of their logic to be masking passively broadcast technosignatures [22]. Humankind has done a rather good job of cleaning up CFCs, but other atmospheric pollutants anomalous in oxidizing atmospheres remain, notably ammonia, anthropogenic Non-Methane Volatile Organic Compounds (NMVOCs) and heavy metal particulates derived from smelting and refining practices. Additionally, waste heat from power generation and consumption may in time prove detectable to ETs, alongside the spectral signatures of city lights. Assuming precautionary malevolence, all these things are immense liabilities; shall we curtail the practices that create them to drive these pollutants below any conceivable detection limit? Take steps no nation or environmental group is prepared to make to reduce humanity's spectral footprint, in a world unable to swiftly deal with anthropogenic climate change? The authors' feelings on environmental and light pollution aside, these conclusions are absurd and unworkable, and no mandate exists to enact them; it is in this wise that passivist arguments reduce to primitivist critiques, again a strain of thought as dangerous to SETI as to METI.

Passivist arguments also entail fear that Earth, a smallish planet squarely in the LWHZ and replete with resources 'ripe for the taking,' will be subject not just to SETI, but to general astrobiological scrutiny. Indeed, Earth is far enough from its parent star to be directly imaged by coronagraph-equipped, next-generation telescopes, making this an easy task [31]. Ought we go still further and dedicate our energies to pumping out reducing antibiosignatures like $H_2$ and CO [42] to ward off the evil aliens? Shall we smother the biosphere to hide chlorophyll's ubiquitous spectral fingerprint? Never mind the consequences, it could be said, 'we're holding the fate of the Earth in our hands!'

There is no end to this slippery slope as there is no escaping Earth's status as a putative subject of SETI. Passivists attempt to square this circle by ignoring the detectability of other technosignatures and arbitrarily propose policy options that sufficiently assuage their worries while sacrificing as little comfort as possible. The scientific community need not credit the alarmist projections of these master hypocrites, and policymaking based on their fears also would run SETI's funding and credibility firmly into the ground.

## 2.5 Securing SETI's Funding and Success

Passivist arguments present a serious threat to the larger SETI project's continued funding and support. Since 1993, American SETI endeavors have been forced to derive funding from transient, often philanthropic sources; Yuri Milner's *Breakthrough Listen* initiative has been a massive boon to the field, but there is no guarantee that such private charity will continue. Passive SETI's likelihood of success at any given time is slim; why pour capital into an effort that is virtually guaranteed not to succeed this quarter, or the next, or the one after? Here opposition to METI is no help. METI marginally increases the likelihood of discovery per unit time, thus (however slightly) increasing the likelihood of returns on investment. Again, the SETI Paradox rears its head: what is the point of SETI if, as passivists argue, signaling is too risky? If donors can complete this defeatist syllogism, the SETI project will never secure continuous funding from private sources.

To restore sustainable funding protected by law, SETI's credibility in the eyes of the public and elected politicians needs to be rehabilitated. This may prove difficult, thanks to a phenomenon known as the 'giggle factor' [23]. Talk of 'aliens' makes people laugh, notably Congresspeople and other funding entities. There is no reason to search exclusively for biosignatures when technosignatures may be easier to find [23], but Congress and other government entities *will not take SETI seriously.* The search for biosignatures has thankfully recovered from this laughingstock status, but SETI has yet to escape it. As a result, technosignatures are often considered separate and apart from (or just as a footnote to) serious astrobiological discussion (e.g. [3,4]), when the distinction between them is far more apparent than real. Gertz urges the National Academy of Sciences to release a study in favor of funding SETI, as its recommendations are often considered authoritative by Congress [16]. One sincerely hopes such an effort would succeed, and there have been whispers of change in the wind – but do not positively expect it.

Too often the laughter at SETI projects is not entirely humorous, and it is passivists' fault that this is true. Passivists wish to find ET asymmetrically; they want to spy, not to contact. While all SETI scientists certainly share the view that SETI advances knowledge both of the cosmos and humanity's place in it, the antipathy and melodrama with which passivists describe mutual contact is not lost on the public. Gertz compares ET detection to catastrophic earthquakes and asteroid impacts, and raises the spectre of interstellar attack [15,16]. Notionally serious scientists fret in print about alien invasion [17]. Hawking compares symmetric contact to imperialism and insinuates humanity is not ready for discovery anyway, two things which he cannot claim to know [19]. They oppose METI despite it being a methodological accelerant, and worry that if discovery does happen, Men in Black will appear to hijack and censor this monumental event [16].

It is well-understood that fear, once spread, can provoke irrational behavior in this context [41]; it was not for nothing that Orson Welles' 1938 radio dramatization of *War of the Worlds* excited some two million people to panic in war-jittery America [43]. Passivists' frightened declarations provoke both honest fear of and cynical derision toward SETI from the public and their elected politicians, who promptly refuse to fund it. Gertz finds this disappointment perplexing, then rather dangerously suggests that SETI eschew funding from the National Aeronautics and Space Administration (NASA) and National Science Foundation (NSF) and instead look to the Department of Defense, explicitly recommending that SETI be treated as a problem of military aggression [16]. Gertz may think himself clever, but this is a Faustian bargain the SETI project simply cannot afford to make.

'The fact that SETI is serious hardly requires defense' [16]. Our present conundrum is the natural outcome of stoking fears of alien invasion, and the dread of ETs as aggressors permeates our culture [23]. This fear of contact, and even of discovery, feeds back into the tribalistic mentality that fuels popular but hysterical films like *Independence Day.* Again, *Star Trek* is no model either; what is needed is a middle path, free of both unrealistic fears and aspirations, that treats METI as what it is – an organic and necessary extension to the SETI project.

## 3.0 Conclusion

Crucial to that middle path is the admission that contact can proceed in a great diversity of ways; the spans of thinkable neutral, positive, and negative outcomes are equal [22]. Even so, one can dismiss contrived notions like the zoo hypothesis, and recognize that METI does not uniquely increase risk even should our cosmic correspondents be 'hostile super-civilizations.' Offered below are policy options that aim to enhance METI's scientific and social standing, in the hopes of maintaining and enriching the financial integrity and public acceptance of the larger SETI project.

### 3.1 Policy Recommendations

Excluding the two *Voyager* probes' Golden Records, interstellar messages to date have been sent using mutually unintelligible encoding schemes and without international consultation, proper feasibility studies, ethical concern for miscommunication or even peer review [13,14,44], a suite of facts that belies metiist's past claims to complete scientific rigor. Indeed, there are no guidelines for METI prior to detection of a reply. Earth is detectable to ETs via its continuum of biosignatures, passive technosignatures, and METI; however, METI represents *the only component of Earth's astrobiological signaling that is entirely under our control.* Those who transmit (or authorize transmission) have complete command over the content, strength and direction of the signal. It is outside the scope of this paper to decide who *speaks* for the Earth (as opposed to other forms of signaling), but that question is not to be written off by those maverick metiists who claim it is their right to get a word in before everyone else. Here METI reduces to a commons problem wherein the actions of individuals may decide the course of history for all concerned, even if the risk to humankind is slight [24]. As such, those who act on their own may be 'engaged in unauthorized… diplomacy' [16]; it follows therefore that we ought to at least be careful about what we transmit, and discourage individuals from doing so by themselves. A more specific and workable set of guidelines are needed to better enshrine, establish, and standardize METI methodology, and here is where well-intentioned passivists and metiists agree. As it is impossible to prevent unauthorized transmissions, METI guidelines must be written charitably and reasonably, *so that they will be followed.*

Future transmissions' feasibility ought to be thoroughly studied in the style of other scientific ventures in outer space: both specifying all parameters of the message and the instrument used to transmit it, and assessing the message's detectability as a function of ET's technical capacity and receiver attributes. Additionally, this information must be free and accessible to the public; an 'online database containing such descriptions in a standard uniform format' is in order [13]. Great care must be taken to ensure that any room for miscommunication is negligible, that messages not provoke disgust in ETs, and to draw upon a broad consensus of human cultures and locations (and animal communications studies) in assessing a messages' comprehensibility and content [17,44]. The specifics of such best-practices are the prerogative of the extended SETI community to decide and debate internally before any thought can be made of making recommendations about transmission to policymakers.

Following this messaging standardization, a concerted plan for sending transmissions to specific targets and receiving return messages must also be developed. Choosing targets will doubtless base itself on the astrobiological scrutiny of potentially habitable exoplanets detailed herein, alongside nearby and/or anomalous stars (e.g. Boyajian's Star). Receiving messages will entail regularly searching a small area of sky around the target after the minimum time for transmission and return (given by twice the target's distance in light-years). This will not be an undue burden on other targeted SETI searches or all-sky surveys, unless something is detected, in which case rapid, confirmatory follow-up observations by other telescopes will be necessary. Humankind has time, in the intervening decades between the sending and receiving of messages, to construct and engineer the facilities needed for such a fantastic enterprise both on and off Earth. Thankfully, such instruments are being designed and built today. Additionally, in line with proposed guidelines for passive SETI, it *must* be stipulated that any information regarding return messages be made free and open to the public, to better discourage the knowledge and/or content of received messages from becoming a flashpoint for international conflict [45].

### 3.2 Toward a Cautious Optimism

A change in the zeitgeist is needed; the SETI community must trade the arbitrary and self-contradictory pessimism of passivists for the cosmopolitan hope of escaping the SETI Paradox and participating in an interstellar community. Additionally, metiists ought to drop contrived and unpolished arguments in favor of their position and instead embrace METI's place along Earth's continuum of astrobiological signaling. SETI scientists must work to legitimize SETI in the hearts and minds of both politicians and the public, and the first step towards that is leaving speculation about the outcome of contact behind. To boost the likelihood of discovery and maintain passive SETI's funding, we should exercise every technique available to us – in the hopes of someday making SETI the planetary-scale endeavor that such a rich and informative enterprise deserves to be.

Such a cautious optimism about the SETI project enjoins that the darkness of space is nothing to fear and proposes to send a flare into the void to illuminate the path forward. Detractors of METI are afraid of the dark, yet ought be more afraid of their spectral shadows; moreover, following their fearful reasoning to its logical conclusions results in absurd policy recommendations. If we care for the Search for Extraterrestrial Intelligence at all, we must integrate an active approach – to better ensure both its continuity and eventual success.